\documentclass[aps, prd, preprintnumbers, floatfix, nofootinbib, 12pt]{revtex4-2}

\usepackage[dvips]{graphics}
\usepackage{amsmath,amsfonts,amssymb,amsthm}
\usepackage{bm}
\usepackage{graphicx}
\usepackage{tabularx}
\usepackage{slashed}

\usepackage[usenames,dvipsnames]{color}
  \definecolor{hgreen}{rgb}{0,.5,0}
  \definecolor{hred}{rgb}{.5,0,0}
  \definecolor{hblue}{rgb}{0,0,.5}
\usepackage[colorlinks=true,linkcolor=hblue,citecolor=hgreen,filecolor=hblue,urlcolor=hred]{hyperref}

\usepackage[english]{babel}
\usepackage[utf8]{inputenc}

\allowdisplaybreaks

\begin{document}
%%%%%%%%%%%%%%%%%%%%%%%%%%%%%%%%%%%%%%%%%%%%%%%%%%%%%%
\title{New Strategies for New Physics Search with \texorpdfstring{\boldmath $\Lambda_b \to \Lambda \nu \bar\nu$}{Lambda b -> Lambda nu nu-bar} Decays}
%%%%%%%%%%%%%%%%%%%%%%%%%%%%%%%%%%%%%%%%%%%%%%%%%%%%%%

\author{Wolfgang~Altmannshofer}
\email{waltmann@ucsc.edu}
\affiliation{Department of Physics, University of California Santa Cruz, and
Santa Cruz Institute for Particle Physics, 1156 High St., Santa Cruz, CA 95064, USA}

\author{Sri Aditya Gadam}
\email{sgadam@ucsc.edu}
\affiliation{Department of Physics, University of California Santa Cruz, and
Santa Cruz Institute for Particle Physics, 1156 High St., Santa Cruz, CA 95064, USA}

\author{Kevin~Toner}
\email{ktoner@ucsc.edu}
\affiliation{Department of Physics, University of California Santa Cruz, and
Santa Cruz Institute for Particle Physics, 1156 High St., Santa Cruz, CA 95064, USA}

%%%%%%%%%%%%%%%%%%
\begin{abstract}
We examine the new physics sensitivity of the rare decay $\Lambda_b \to \Lambda \nu\bar\nu$ which can be accessible at future $Z$-pole machines like FCC-ee and CEPC. We find that the longitudinal polarization of $\Lambda_b$ baryons produced in $Z$ decays introduces a novel observable, the forward-backward asymmetry $A_\text{FB}^\uparrow$ in the angle between the outgoing $\Lambda$ momentum and the $\Lambda_b$ spin. We provide Standard Model predictions for the $\Lambda_b \to \Lambda \nu\bar\nu$ branching ratio and $A_\text{FB}^\uparrow$, and show that future precision measurements of these observables are complementary and probe new physics scales comparable to other $b \to s \nu\bar\nu$ and $b \to s \ell^+ \ell^-$ processes. We also show that the zero crossing of the forward-backward asymmetry offers a robust test of form factor calculations independent of new physics.
\end{abstract}
%%%%%%%%%%%%%%%%%%

\maketitle
%\tableofcontents

%%%%%%%%%%%%%%%%%%%%%%%%%
\section{Introduction}
\label{sec:intro}
%%%%%%%%%%%%%%%%%%%%%%%%%

Rare $b$ decays are highly suppressed in the Standard Model and therefore exceptionally sensitive to new physics at very high energy scales. They garner much effort and attention from the theoretical and experimental communities alike~\cite{Blake:2016olu, Altmannshofer:2022hfs}. 

Precise experimental measurements and accurate SM predictions are essential to fully harness the potential of rare $b$ decays as probes of new physics.
Experimental results from LHCb, ATLAS, and CMS exist for a multitude of rare semi-leptonic $B$ meson decays including $B \to K \mu^+ \mu^-$, $B \to K^* \mu^+ \mu^-$ and $B_s \to \phi \mu^+ \mu^-$. Measured observables include differential branching ratios~\cite{LHCb:2014cxe, LHCb:2016ykl,  LHCb:2021zwz}, angular distributions~\cite{LHCb:2015svh, CMS:2017rzx, ATLAS:2018gqc, LHCb:2020lmf, LHCb:2020gog, LHCb:2021xxq, LHCb:2023gel, LHCb:2023gpo, LHCb:2024onj, CMS:2024atz}, and lepton flavor universality ratios~\cite{LHCb:2021lvy, LHCb:2022qnv, LHCb:2022vje, CMS:2024syx, LHCb:2024rto}. To maximize the potential of rare decay observables in probing new physics, global fits incorporating all relevant rare $b$ decay data are frequently performed, for recent results see~\cite{Altmannshofer:2021qrr, SinghChundawat:2022zdf, Ciuchini:2022wbq, Greljo:2022jac, Alguero:2023jeh, Wen:2023pfq, Altmannshofer:2023uci, Guadagnoli:2023ddc, Hurth:2023jwr, Bordone:2024hui}. Intriguingly, there are several measured branching ratios and angular observables that do not agree well with SM predictions and that fairly consistently point to new physics. The interpretation of the observed deviations in terms of new physics relies critically on the robust control of hadronic effects in $b \to s \ell^+ \ell^-$ decays, parameterized by local and non-local form factors~\cite{Horgan:2013pva, Bouchard:2013eph, Horgan:2015vla, Bailey:2015dka, Gubernari:2018wyi, Gubernari:2022hxn, Gubernari:2023puw, Aliev:2010uy}.

In this context, the rare decays based on the $b \to s \nu\bar\nu$ transition emerge as an interesting complementary probe of new physics~\cite{Altmannshofer:2009ma, Buras:2014fpa, Browder:2021hbl, Bause:2021cna, Becirevic:2023aov, Rajeev:2021ntt}. The di-neutrino modes are theoretically cleaner than their charged-lepton counterparts, as they are not affected by the non-local hadronic effects (e.g. the charm loops). All hadronic physics in the exclusive $b \to s \nu\bar{\nu}$ decays is encapsulated by local form factors. The $b \to s \nu\bar{\nu}$ modes play a dual role in probing new physics: they test heavy new physics through model-independent four-fermion contact interactions (potentially linked to $b \to s \ell^+ \ell^-$ transitions via $SU(2)_L$ symmetry) and provide a window into light dark sectors, as decays like $b \to s X$---where $X$ is a neutral, invisibly decaying, or long-lived light particle---mimic the missing-energy signature of di-neutrino processes.

On the experimental side, BaBar and Belle established upper limits on the $B \to K \nu\bar\nu$ and $B \to K^* \nu\bar \nu$ branching ratios a factor of few above the SM predictions~\cite{BaBar:2010oqg, Belle:2013tnz, BaBar:2013npw, Belle:2017oht}, while Belle~II recently found first evidence for $B^+ \to K^+ \nu\bar\nu$~\cite{Belle-II:2023esi}. The measured branching ratio is $\sim 2.7\sigma$ above the SM expectation, which sparked renewed interest in the $b \to s \nu\bar\nu$ decays as probes of new physics~\cite{Bause:2023mfe, Allwicher:2023xba, Felkl:2023ayn, Dreiner:2023cms, He:2023bnk, Berezhnoy:2023rxx, Datta:2023iln, Altmannshofer:2023hkn, McKeen:2023uzo, Fridell:2023ssf, Ho:2024cwk, Chen:2024jlj, Gabrielli:2024wys, Hou:2024vyw, He:2024iju, Bolton:2024egx, Marzocca:2024hua, Buras:2024ewl, Allwicher:2024ncl, Altmannshofer:2024kxb, Hu:2024mgf}. Belle~II is expected to measure the $B^+ \to K^+ \nu\bar\nu$ and $B^0 \to K^{*0} \nu\bar \nu$ branching ratios with $\sim 10\%$ precision~\cite{Belle-II:2018jsg}. Future $e^+ e^-$ machines running on the $Z$-pole like FCC-ee and CEPC~\cite{FCC:2018evy, Bernardi:2022hny, CEPCStudyGroup:2018ghi, CEPCPhysicsStudyGroup:2022uwl} have the potential to further improve this precision and have the unique capability to observe the $B_s\to \phi \nu\bar\nu$ and $\Lambda_b \to \Lambda \nu\bar\nu$ decays~\cite{Li:2022tov, Amhis:2023mpj}. Combining the insights from the full range of di-neutrino modes---including pseudoscalar-to-pseudoscalar transitions ($B \to K\nu \bar{\nu}$), pseudoscalar-to-vector transitions ($B \to K^*\nu \bar{\nu}$, $B_s \to \phi\nu \bar{\nu}$), and fermion-to-fermion transitions ($\Lambda_b \to \Lambda\nu \bar{\nu}$)---offers a powerful approach to probing new physics.

The decays of the $\Lambda_b$ baryon are particularly interesting, as the $\Lambda_b$ can be polarized and thus provide novel observables that are not accessible in the meson decays. Here, we will be especially interested in the fact that $\Lambda_b$ baryons that are produced in $Z$ decays are longitudinally polarized~\cite{Mannel:1991bs, Falk:1993rf, ALEPH:1995aqx, OPAL:1998wmk, DELPHI:1999hkl}. The goal of our work is to provide a comprehensive study of the decay $\Lambda_b \to \Lambda \nu\bar\nu$ of longitudinally polarized $\Lambda_b$ baryons (see~\cite{Chen:2000mr, Aliev:2007rm, Sirvanli:2007yq, Hiller:2021zth, Das:2023kch, Amhis:2023mpj}, for related work on $\Lambda_b \to \Lambda \nu\bar\nu$ decays). The kinematics of the decay is characterized by the energy of the $\Lambda$ and the angle between its momentum and the spin of the $\Lambda_b$. We find that the corresponding angular decay distribution depends on a novel observable, a forward-backward asymmetry $A_\text{FB}^\uparrow$, that is proportional to the longitudinal polarization of the $\Lambda_b$. We work out state-of-the-art SM predictions for the integrated branching ratio, $\text{BR}(\Lambda_b \to \Lambda \nu\bar\nu)$, and the forward-backward asymmetry, $A_\text{FB}^\uparrow$, and explore their sensitivity to new physics, in particular to the new physics' chirality structure. See~\cite{Hiller:2001zj} for similar ideas in the context of the $\Lambda_b \to \Lambda \gamma$ decay.

The corresponding decays with charged leptons in the final state, $\Lambda_b \to \Lambda \ell^+\ell^-$, have been observed at LHCb~\cite{LHCb:2015tgy, LHCb:2018jna} and are extensively discussed in the literature~\cite{Detmold:2012vy, Boer:2014kda, Mott:2015zma, Detmold:2016pkz, Blake:2017une, Roy:2017dum, Blake:2019guk, Blake:2022vfl}. It would be interesting to extend our work, and to study the decays of longitudinally polarized $\Lambda_b \to \Lambda \ell^+\ell^-$, at future $Z$-pole machines. Note that in our paper, we focus on the decay $\Lambda_b \to \Lambda \nu \bar\nu$, where $\Lambda$ refers to the weakly decaying $\Lambda$ baryon ground state. An analogous analysis could also be performed for the $\Lambda_b \to \Lambda(1520) \nu \bar\nu$ decay or other higher excited $\Lambda$ baryon states. See~\cite{Meinel:2020owd, Bordone:2021bop, Meinel:2021mdj, Amhis:2022vcd} for the relevant transition form factors, and~\cite{Descotes-Genon:2019dbw, Das:2020cpv, Amhis:2020phx, Li:2022nim, Beck:2022spd, LHCb:2017slr, LHCb:2019efc, LHCb:2023ptw, LHCb:2024ick} for related work on $\Lambda_b \to \Lambda(1520) \ell^+ \ell^-$ decays.

This paper is organized as follows: In section~\ref{sec:H_eff}, we introduce the effective Hamiltonian that describes the $b \to s \nu\bar\nu$ decay in the SM and in models with heavy new physics and define the relevant $\Lambda_b \to \Lambda$ form factors. We discuss in detail the uncertainties from the short distance SM contributions and the CKM input. In section~\ref{sec:dBR}, we present our results for the differential $\Lambda_b \to \Lambda \nu\bar\nu$ branching ratio in the presence of longitudinal $\Lambda_b$ polarization and introduce the forward-backward asymmetry $A_\text{FB}^\uparrow$. We give SM predictions for the branching ratio and the forward-backward asymmetry and discuss qualitatively the impact of new physics on the observables. In section~\ref{sec:NP} we determine the new physics sensitivity of precision measurements of the $\Lambda_b \to \Lambda \nu\bar\nu$ observables. Finally, in section~\ref{sec:lab_frame}, we discuss the imprint of the forward-backward asymmetry on the $\Lambda$ energy distribution in the lab frame. This distribution might provide an alternative method to test new physics with $\Lambda_b \to \Lambda \nu\bar\nu$. We conclude in section~\ref{sec:conclusions}. Details about our implementation of the $\Lambda_b \to \Lambda$ form factors are collected in appendix~\ref{app:ff}. 

%%%%%%%%%%%%%%%%%%%%%%%%%
\section{Effective Hamiltonian and Hadronic Matrix Elements}
\label{sec:H_eff}
%%%%%%%%%%%%%%%%%%%%%%%%%

The effective Hamiltonian that underlies the description of the $\Lambda_b \to \Lambda \nu \bar \nu$ decay is
\begin{equation} \label{eq:H_eff}
\mathcal H_\text{eff} = -\frac{4 G_F}{\sqrt{2}} \frac{\alpha}{4 \pi} V_{ts}^* V_{tb} \Big( C_L O_L + C_R O_R \Big) ~+~ \text{h.c.} ~,
\end{equation}
with the four-fermion operators
\begin{equation}
O_L = (\bar s \gamma^\mu P_L b)(\bar \nu \gamma_\mu (1 - \gamma_5) \nu) ~,\qquad O_R = (\bar s \gamma^\mu P_R b)(\bar \nu \gamma_\mu (1 - \gamma_5) \nu) ~.
\end{equation}
In the SM, the Wilson coefficients are $C_L^\text{SM} = - X(x_t)/s_W^2$ where $s_W = \sin\theta_W$ is the sine of the weak mixing angle, and $C_R^\text{SM} \simeq 0$. The loop function $X(x_t)$ depends on the ratio of the top quark $\overline{\text{MS}}$ mass and the $W$ mass $x_t = m_t^2(m_t)/m_W^2$. Its most recent determination can be found in~\cite{Brod:2010hi, Brod:2021hsj}.
The uncertainties in the $W$ mass and the sine of the weak mixing angle $s_W$ can be neglected when evaluating the SM Wilson coefficient, but the uncertainty in the top mass is not entirely negligible.

To obtain a value for $C_L^\text{SM}$, we re-scale the $X(x_t)$ function given in~\cite{Brod:2021hsj} to take into account the latest experimental results on the top mass. We translate the top quark pole mass from cross-section measurements quoted by the PDG, $m_t = (172.5 \pm 0.7)$\,GeV~\cite{ParticleDataGroup:2022pth}, into the top quark $\overline{\text{MS}}$ mass at 3-loop QCD accuracy using \texttt{RunDec}~\cite{Herren:2017osy} and find $m_t(m_t) = 162.92 \pm 0.72$\,GeV. Using $m_W = 80.377$\,GeV and $s_W^2 = 0.2314$~\cite{ParticleDataGroup:2022pth} we arrive at
\begin{equation}
C_L^\text{SM} = -6.322 \pm 0.031\Big|_{m_t} \pm 0.074\Big|_\text{QCD} \pm 0.009\Big|_\text{EW} ~,
\end{equation}
where the three uncertainties are due to the top mass, higher-order QCD corrections, and higher-order EW corrections, respectively. The value of the electromagnetic coupling $\alpha$ that enters the effective Hamiltonian is the running $\alpha$ at the electroweak scale, $\alpha^{-1} \simeq 127.95$ with negligible uncertainty.

We determine the relevant CKM matrix elements that enter the effective Hamiltonian from input given by the PDG. In particular, we use~\cite{ParticleDataGroup:2022pth}
\begin{equation} \label{eq:CKM_input}
    |V_{cb}| = (40.8 \pm 1.4)\times 10^{-3}~,~~~ |V_{ub}| = (3.82 \pm 0.20)\times 10^{-3}~,~~~ \gamma = 65.9^\circ \pm 3.5^\circ~.
\end{equation}
The values for $|V_{cb}|$ and $|V_{ub}|$ are conservative averages of determinations using inclusive and exclusive tree-level $B$ decays.
For the sine of the Cabibbo angle, we use $\lambda \simeq 0.225$~\cite{ParticleDataGroup:2022pth}, neglecting its tiny uncertainty.
This results in
\begin{equation}
|V_{ts}^* V_{tb}|^2  = \Big(1.609 \pm 0.109\Big|_{V_{cb}} \pm 0.001\Big|_{V_{ub}} \pm 0.004\Big|_\gamma \Big) \times 10^{-3} ~,
\end{equation}
with the uncertainty entirely dominated by $|V_{cb}|$. 

The hadronic matrix elements relevant for the $\Lambda_b \to \Lambda \nu\bar\nu$ decay can be parameterized by form factors in the following way~\cite{Feldmann:2011xf, Detmold:2016pkz, Blake:2022vfl}
\begin{multline}
\langle \Lambda | \bar s \gamma^\mu b |\Lambda_b \rangle  = \bar u_\Lambda \Bigg[ f_t^V(q^2) (m_{\Lambda_b} - m_\Lambda) \frac{q^\mu}{q^2} + f_\perp^V(q^2) \left( \gamma^\mu - \frac{2 (m_\Lambda P^\mu + m_{\Lambda_b} p^\mu)}{(m_{\Lambda_b} + m_\Lambda)^2 - q^2} \right)  \\
+ f_0^V(q^2) \frac{m_{\Lambda_b} + m_\Lambda}{(m_{\Lambda_b} + m_\Lambda)^2 - q^2} \left( P^\mu + p^\mu - (m_{\Lambda_b}^2 - m_\Lambda^2) \frac{q^\mu}{q^2} \right) \Bigg] u_{\Lambda_b}~,
\end{multline}
\begin{multline}
\langle \Lambda | \bar s \gamma^\mu \gamma_5 b |\Lambda_b \rangle  = - \bar u_\Lambda \gamma_5 \Bigg[ f_t^A(q^2) (m_{\Lambda_b} + m_\Lambda) \frac{q^\mu}{q^2} + f_\perp^A(q^2) \left( \gamma^\mu + \frac{2 (m_\Lambda P^\mu - m_{\Lambda_b} p^\mu)}{(m_{\Lambda_b} - m_\Lambda)^2 - q^2} \right)  \\
+ f_0^A(q^2) \frac{m_{\Lambda_b} - m_\Lambda}{(m_{\Lambda_b} - m_\Lambda)^2 - q^2} \left( P^\mu + p^\mu - (m_{\Lambda_b}^2 - m_\Lambda^2) \frac{q^\mu}{q^2} \right) \Bigg] u_{\Lambda_b}~,
\end{multline}
where $P$ and $p$ are the momenta of the $\Lambda_b$ and $\Lambda$, respectively, and $m_{\Lambda_b}$ and $m_\Lambda$ are their masses. The form factors $f_{t,0,\perp}^{V,A}$ are functions of $q^2 = (P - p)^2$, the di-neutrino invariant mass squared. 
For our numerical analysis, we use the $\Lambda_b \to \Lambda$ form factors from~\cite{Detmold:2016pkz} (see also~\cite{Blake:2022vfl}). More details are provided in appendix~\ref{app:ff}.

%%%%%%%%%%%%%%%%%%%%%%%%%
\section{Differential Branching Ratio in the SM and Beyond}
\label{sec:dBR}
%%%%%%%%%%%%%%%%%%%%%%%%%

With the effective Hamiltonian introduced in the previous section, we can determine the differential branching ratio of the $\Lambda_b \to \Lambda \nu\bar\nu$ decay.
For the decay of a polarized $\Lambda_b$, the kinematics of the visible decay products depends on two independent variables. One convenient choice is $E_\Lambda$, the energy of the $\Lambda$ in the $\Lambda_b$ restframe, and $\theta_\Lambda$, the angle between the $\Lambda$ momentum and the spin quantization axis of the $\Lambda_b$ in its restframe. Alternatively, one may want to use the di-neutrino invariant mass squared $q^2$ and the angle $\theta_\Lambda$, with $q^2$ given in terms of the $\Lambda$ energy by
\begin{equation} \label{eq:q2}
q^2 = m_{\Lambda_b}^2 + m_\Lambda^2 - 2 m_{\Lambda_b} E_\Lambda ~.
\end{equation} 
We find that the double differential branching ratio has a very simple dependence on $\cos\theta_\Lambda$ that can be written as
\begin{equation} \label{eq:dBR_dE_dcos}
 \frac{d\text{BR}(\Lambda_b \to \Lambda \nu\bar\nu)}{dE_\Lambda d\cos\theta_\Lambda} = \frac{d\text{BR}(\Lambda_b \to \Lambda \nu\bar\nu)}{dE_\Lambda} \left( \frac{1}{2} + A_\text{FB}^\uparrow \cos\theta_\Lambda \right) ~.
\end{equation}
The differential branching ratio as a function of the $\Lambda$ energy is given by
\begin{multline} \label{eq:dBR_dE}
 \qquad \frac{d\text{BR}(\Lambda_b \to \Lambda \nu\bar\nu)}{dE_\Lambda} = \tau_{\Lambda_b} \frac{\alpha^2 G_F^2}{32 \pi^5} m_{\Lambda_b}^5 |V_{ts}^* V_{tb}|^2 \frac{E_{\Lambda}^2 }{m_{\Lambda_b}^3} \sqrt{1 - \frac{m_\Lambda^2}{E_\Lambda^2}} \\
 \times 
 \Big( \big| C_L + C_R \big|^2 \mathcal F_V + \big| C_L - C_R \big|^2 \mathcal F_A  \Big) ~, \qquad
\end{multline}
where we summed over all three neutrino flavors, assuming that the Wilson coefficients are neutrino flavor conserving and neutrino flavor universal, as is the case in the SM. The numerical value for the $\Lambda_b$ lifetime that enters the branching ratio is $\tau_{\Lambda_b} = (0.970 \pm 0.009) \times \tau_{B^0} = (1.473 \pm 0.014) \times 10^{-12}$\,s~\cite{ParticleDataGroup:2022pth}.

The quantity $A_\text{FB}^\uparrow$ is a forward-backward asymmetry of the $\Lambda$ with respect to the spin quantization axis of the $\Lambda_b$. We find 
\begin{equation}  \label{eq:A_FB}
A_\text{FB}^\uparrow = \frac{\mathcal P_{\Lambda_b} \big( \big|C_R \big|^2 - \big|C_L \big|^2 \big) \mathcal F_{VA}}{\big| C_L + C_R \big|^2 \mathcal F_V + \big| C_L - C_R \big|^2 \mathcal F_A} ~.
\end{equation}
To simplify notation in the above expressions for the branching ratio and the forward-backward asymmetry, we have introduced the following quadratic functions of the $\Lambda_b \to \Lambda$ form factors
\begin{eqnarray} \label{eq:F_V}
 \mathcal F_V &=& \left( 1 - \frac{m_\Lambda}{E_\Lambda} \right) \left( \left( 1 +x_\Lambda \right)^2 \big(f_0^V(q^2)\big)^2 +  \frac{2q^2}{m_{\Lambda_b}^2} \big(f_\perp^V(q^2)\big)^2 \right) ~, \\  \label{eq:F_A}
 \mathcal F_A &=& \left( 1 + \frac{m_\Lambda}{E_\Lambda} \right) \left( \left( 1 - x_\Lambda\right)^2 \big(f_0^A(q^2)\big)^2  + \frac{2q^2}{m_{\Lambda_b}^2} \big(f_\perp^A(q^2)\big)^2 \right)~, \\ \label{eq:F_VA}
 \mathcal F_{VA} &=& \sqrt{1 - \frac{m_\Lambda^2}{E_\Lambda^2}} \left( \left( 1 - x_\Lambda^2 \right) f_0^V(q^2) f_0^A(q^2)  - \frac{2q^2}{m_{\Lambda_b}^2}  f_\perp^V(q^2) f_\perp^A(q^2)  \right)~.
\end{eqnarray}
All form factors in the above expressions depend on the di-neutrino invariant mass squared $q^2$ as introduced already in~\eqref{eq:q2}. For later convenience, we have introduced the ratio of $\Lambda$ mass and $\Lambda_b$ mass $x_\Lambda = m_\Lambda/m_{\Lambda_b}$. 

The quantity $\mathcal P_{\Lambda_b}$ that enters the forward-backward asymmetry is the polarization asymmetry of the $\Lambda_b$, i.e. it corresponds to the difference of $\Lambda_b$ baryons with spin-up and spin-down
\begin{equation}
 \mathcal P_{\Lambda_b} = \frac{N_{\Lambda_b}^\uparrow - N_{\Lambda_b}^\downarrow}{N_{\Lambda_b}^\uparrow + N_{\Lambda_b}^\downarrow} ~.
\end{equation}
The natural choice for the spin quantization axis is the direction of the $\Lambda_b$ momentum in the lab frame. In that case, $\mathcal P_{\Lambda_b}$ corresponds to the longitudinal polarization fraction of the $\Lambda_b$. On the $Z$ pole, this polarization can be measured using semi-leptonic $\Lambda_b$ decays~\cite{Mannel:1991bs, Mele:1992kh, Bonvicini:1994mr, Diaconu:1995mp}. Measurements at LEP found
\begin{equation}
\mathcal{P}_{\Lambda_b} = \begin{cases} &- 0.23^{+0.24}_{-0.20} {}^{+0.08}_{-0.07} ~, \qquad \text{ALEPH~\cite{ALEPH:1995aqx}} ~, \\
&- 0.49 ^{+0.32}_{-0.30} \pm 0.17 ~, \quad \text{DELPHI~\cite{DELPHI:1999hkl}} ~, \\
&- 0.56 ^{+0.20}_{-0.13} \pm 0.09 ~, \quad \text{OPAL~\cite{OPAL:1998wmk}} ~, \end{cases}
\end{equation}
where the first uncertainty is due to statistics and the second due to systematics.
A naive weighted average of the LEP measurements that neglects possible correlations gives
\begin{equation} \label{eq:P_Lambdab_WA}
\mathcal P_{\Lambda_b} =  - 0.40 \pm 0.14 ~.
\end{equation}
We expect that a future $Z$ pole machine should be able to measure $\mathcal P_{\Lambda_b} $ with percent level accuracy. In fact, the statistical uncertainty at FCC-ee or CEPC should be negligible, while systematic uncertainties from e.g. the lepton energy and missing energy resolution should be improved significantly compared to LEP.\footnote{Note that even in the challenging environment of the LHC, measurements of the (transversal) $\Lambda_b$ polarization with an uncertainty of few percent have been performed~\cite{CMS:2018wjk, LHCb:2020iux}}

%%%%%%%%%%%%%%%%%%%%%%%%%%%%%%%%%%%%%%%%%%%%%%%
\subsection{Standard Model Predictions}
%%%%%%%%%%%%%%%%%%%%%%%%%%%%%%%%%%%%%%%%%%%%%%%

Before discussing the new physics sensitivity of the $\Lambda_b \to \Lambda \nu\bar\nu$ decay, we provide the SM predictions for the branching ratio and the forward-backward asymmetry.

The SM expression for the integrated branching ratio can be found from~\eqref{eq:dBR_dE}, setting the right-handed Wilson coefficient to zero
\begin{equation} \label{eq:BR}
 \text{BR}(\Lambda_b \to \Lambda \nu\bar\nu)_\text{SM} = \tau_{\Lambda_b} \frac{\alpha^2 G_F^2}{32 \pi^5} m_{\Lambda_b}^5 |V_{ts}^* V_{tb}|^2 \big| C_L^\text{SM} \big|^2 \int_{E_\Lambda^\text{min}}^{E_\Lambda^\text{max}} \frac{dE_\Lambda E_\Lambda^2}{m_{\Lambda_b}^3}  \sqrt{1 - \frac{m_\Lambda^2}{E_\Lambda^2}} ~\big( \mathcal F_V + \mathcal F_A \big) ~,
\end{equation}
where $\mathcal F_V$ and $\mathcal F_A$ were already introduced in eqs.~\eqref{eq:F_V} and \eqref{eq:F_A}, and the integration boundaries are given by
\begin{equation}
  E_\Lambda^\text{min} = m_\Lambda ~,\qquad E_\Lambda^\text{max} = \frac{m_{\Lambda_b}}{2} \left( 1 + x_\Lambda^2 \right) ~.
\end{equation} 
We stress that the branching ratio is independent of the $\Lambda_b$ polarization $\mathcal P_{\Lambda_b}$.

%%%%%%%%%%%%%%%%%%%%%%%%%%%%%%%%%%%%%%
\begin{figure}[tb]
\centering
\includegraphics[width=0.72\textwidth]{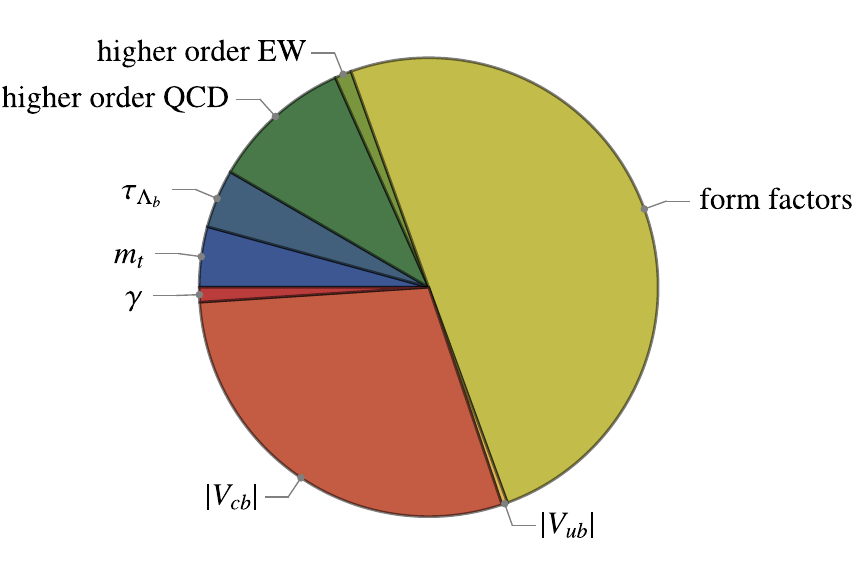}
\caption{Pie chart that shows the most relevant sources of uncertainty entering the SM prediction of the $\Lambda_b \to \Lambda \nu \bar \nu$ branching ratio.}
\label{fig:pie}
\end{figure}
%%%%%%%%%%%%%%%%%%%%%%%%%%%%%%%%%%%%%%

Using the numerical input discussed previously, we arrive at the following numerical SM prediction for the integrated branching ratio 
\begin{equation} \label{eq:SM_prediction}
\text{BR}(\Lambda_b \to \Lambda \nu\bar\nu)_\text{SM} = (7.71  \pm 1.06) \times 10^{-6}  ~,
\end{equation}
which has an uncertainty of approximately $14\%$. In figure~\ref{fig:pie}, we show a pie chart of the various sources of uncertainty that enter our SM prediction. The bulk of the uncertainty is shared by the form factors and the CKM input, with the form factors being dominant. The other sources of uncertainty are subdominant
\begin{multline}
 \delta\text{BR}(\Lambda_b \to \Lambda \nu\bar\nu)_\text{SM} \times 10^6 = \pm 0.90 \Big|_\text{ff} \pm 0.53 \Big|_\text{CKM} \pm 0.18 \Big|_\text{QCD} \\ \pm 0.08 \Big|_{m_t} \pm 0.07 \Big|_{\tau_{\Lambda_b}} \pm 0.02 \Big|_\text{EW} = \pm 1.06~,
\end{multline}
where we added all uncertainties in quadrature. 

%%%%%%%%%%%%%%%%%%%%%%%%%%%%%%%%%%%%%%
\begin{figure}[tb]
\centering
\includegraphics[width=0.45\textwidth]{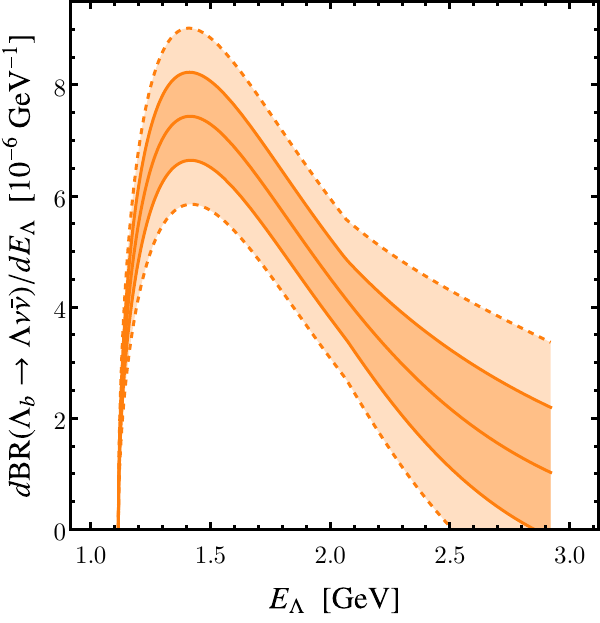} ~~~~
\includegraphics[width=0.46\textwidth]{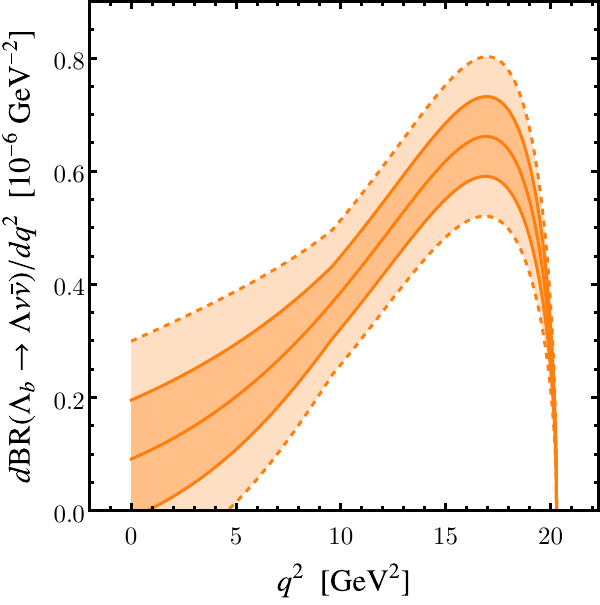}
\caption{The differential branching ratio of $\Lambda_b \to \Lambda \nu \bar \nu$ in the SM as a function of the $\Lambda$ energy in the $\Lambda_b$ rest frame $E_\Lambda$ (left) and the di-neutrino invariant mass squared $q^2$ (right). The relevant phase space boundaries of the decay are $m_\Lambda < E_\Lambda < \frac{m_{\Lambda_b}}{2}(1 + m_\Lambda^2 / m_{\Lambda_b}^2 )$ and $0 < q^2 < (m_{\Lambda_b}- m_\Lambda)^2$. The colored bands correspond to the $1\sigma$ and $2\sigma$ uncertainties.}
\label{fig:dBR_LbtoLnunu}
\end{figure}
%%%%%%%%%%%%%%%%%%%%%%%%%%%%%%%%%%%%%%
\begin{figure}[tb]
\centering
\includegraphics[width=0.45\textwidth]{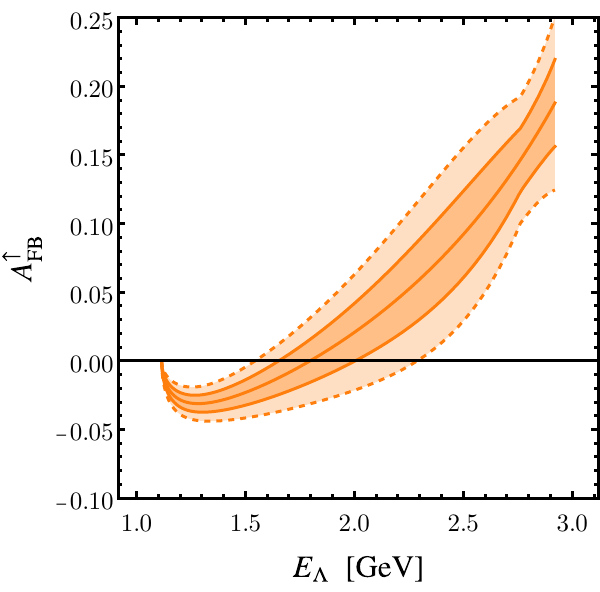} ~~~~
\includegraphics[width=0.46\textwidth]{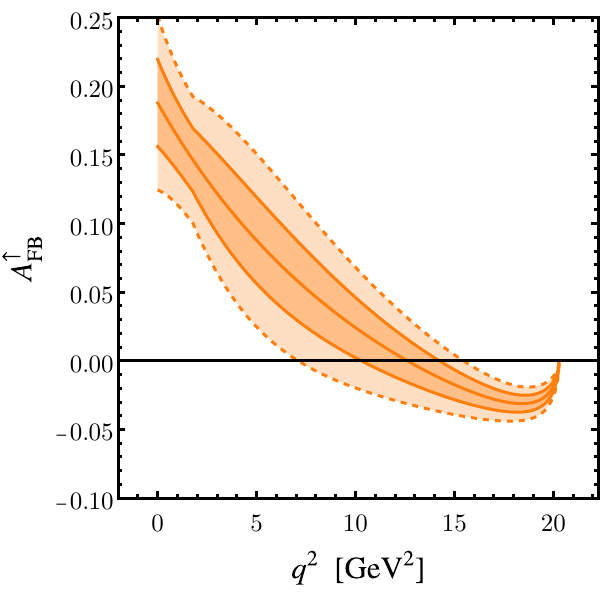}
\caption{The forward-backward asymmetry in $\Lambda_b \to \Lambda \nu \bar \nu$ in the SM as a function of the $\Lambda$ energy in the $\Lambda_b$ rest frame $E_\Lambda$ (left) and the di-neutrino invariant mass squared $q^2$ (right). The relevant phase space boundaries of the decay are $m_\Lambda < E_\Lambda < \frac{m_{\Lambda_b}}{2}(1 + m_\Lambda^2 / m_{\Lambda_b}^2 )$ and $0 < q^2 < (m_{\Lambda_b}- m_\Lambda)^2$. The colored bands correspond to the $1\sigma$ and $2\sigma$ uncertainties. Only uncertainties due to the form factors are shown. The $\Lambda_b$ polarization is set to its central experimental value $\mathcal P_{\Lambda_b} = -0.40$.}
\label{fig:AFB}
\end{figure}
%%%%%%%%%%%%%%%%%%%%%%%%%%%%%%%%%%%%%%

In figure~\ref{fig:dBR_LbtoLnunu}, we show the differential branching ratio~\eqref{eq:dBR_dE} in the SM as a function of the $\Lambda$ energy $E_\Lambda$ (left panel) and the di-neutrino invariant mass squared (right panel). Also these spectra are independent of the $\Lambda_b$ polarization $\mathcal P_{\Lambda_b}$. The uncertainty increases for small $q^2$ or large $E_\Lambda$, respectively, as the uncertainties in the form factors increases in that kinematic regime.

The forward-backward asymmetry in the SM is given by
\begin{equation}
(A_\text{FB}^\uparrow)_\text{SM} = \frac{-\mathcal P_{\Lambda_b} \mathcal F_{VA}}{\mathcal F_V + \mathcal F_A} ~.
\end{equation}
Besides $\mathcal P_{\Lambda_b}$, the only other source of uncertainty are the form factors contained in the functions $\mathcal F_V$, $\mathcal F_A$, and $\mathcal F_{VA}$ in eqs.~\eqref{eq:F_V}, \eqref{eq:F_A}, and \eqref{eq:F_VA}. In figure~\ref{fig:AFB}, we show $(A_\text{FB}^\uparrow)_\text{SM}$ as a function of the $\Lambda$ energy $E_\Lambda$ (left panel) and the di-neutrino invariant mass squared (right panel). For concreteness we set the $\Lambda_b$ polarization to the central value given in eq.~\eqref{eq:P_Lambdab_WA}. The shown uncertainty is from the form factors only. The forward-backward asymmetry vanishes at the kinematic endpoint $E_\Lambda^\text{min} = m_\Lambda$, or correspondingly $q^2_\text{max} = (m_{\Lambda_b} - m_\Lambda)^2$. We find that the forward-backward asymmetry has a zero crossing in the SM, which is determined by the relative size of the form factors $f_0^{V,A}$ and $f_\perp^{V,A}$, cf. equation~\eqref{eq:F_VA}, by the implicit relation
\begin{equation} \label{eq:q2_0}
q^2 = \frac{m_{\Lambda_b}^2}{2} \big( 1 - x_\Lambda^2 \big) \frac{f_0^V(q^2) f_0^A(q^2)}{f_\perp^V(q^2) f_\perp^A(q^2)} ~.
\end{equation}
The corresponding numerical zero crossing values for the $\Lambda$ energy and the di-neutrino invariant mass are
\begin{equation} \label{eq:zero_crossing}
 (E_\Lambda)_0^\text{SM} = ( 1.80 \pm 0.11 )~\text{GeV} ~, \quad (q^2)_0^\text{SM} = (12.6 \pm 1.2)~\text{GeV}^2 ~.
\end{equation}
The uncertainties in these values are entirely due to the form factors.

Integrating over the whole range of $\Lambda$ energies we find in the SM
\begin{equation}
\langle A_\text{FB}^\uparrow \rangle_\text{SM} = -\mathcal P_{\Lambda_b} \times \frac{\int_{E_\Lambda^\text{min}}^{E_\Lambda^\text{max}} dE_\Lambda E_\Lambda^2 \sqrt{1 - \frac{m_\Lambda^2}{E_\Lambda^2}} ~ \mathcal F_{VA}}{\int_{E_\Lambda^\text{min}}^{E_\Lambda^\text{max}} dE_\Lambda E_\Lambda^2 \sqrt{1 - \frac{m_\Lambda^2}{E_\Lambda^2}} ~\big( \mathcal F_V + \mathcal F_A \big) } = -\mathcal P_{\Lambda_b} \times \big( 2.7 \pm 3.4 \big) \times 10^{-2} ~.
\end{equation}
The SM prediction of the integrated asymmetry is fairly small, with the uncertainty entirely due to the form factors. If we instead split the integration region into two parts at the zero crossing of the forward-backward asymmetry, we find
\begin{eqnarray}
\langle A_\text{FB}^\uparrow \rangle_\text{SM}^\text{low} &=& -\mathcal P_{\Lambda_b} \times \big( 13.2 \pm 4.2 \big) \times 10^{-2} ~,\quad 0 < q^2 < 12.6\,\text{GeV}^2 ~, \\
\langle A_\text{FB}^\uparrow \rangle_\text{SM}^\text{high} &=& -\mathcal P_{\Lambda_b} \times \big( -5.3 \pm 1.4 \big) \times 10^{-2} ~,  \quad 12.6\,\text{GeV}^2 < q^2 < q^2_\text{max} ~,
\end{eqnarray}
where ``low'' and ``high'' refer to the $q^2$ regions $0 < q^2 < 12.6\,\text{GeV}^2$ and $12.6\,\text{GeV}^2 < q^2 < q^2_\text{max}$, respectively.  
We find that the uncertainties of the forward-backward asymmetries in the low and high $q^2$ regions are correlated, with a correlation coefficient of $\rho \simeq +52\%$. 

Analogously, we can also obtain SM predictions for the branching ratio in the two different $q^2$ regions
\begin{eqnarray}
\text{BR}(\Lambda_b \to \Lambda \nu\bar\nu)_\text{SM}^\text{low} &=& (3.32  \pm 0.99) \times 10^{-6}  ~,\quad 0 < q^2 < 12.6\,\text{GeV}^2 ~, \\
\text{BR}(\Lambda_b \to \Lambda \nu\bar\nu)_\text{SM}^\text{high} &=& (4.39  \pm 0.47) \times 10^{-6}  ~, \quad 12.6\,\text{GeV}^2 < q^2 < q^2_\text{max} ~,
\end{eqnarray}
with an error correlation of $\rho \simeq +40\%$.
The prediction is less precise at low $q^2$, because of the larger form factor uncertainties in this kinematic regime.

As we will see below, splitting the phase space into the two $q^2$ regions below and above the zero crossing of the forward-backward asymmetry enhances the sensitivity to new physics.

%%%%%%%%%%%%%%%%%%%%%%%%%
\subsection{Impact of Heavy New Physics}
%%%%%%%%%%%%%%%%%%%%%%%%%

In this paper we focus on the impact of heavy new physics on the $\Lambda_b \to \Lambda \nu \bar\nu$ decay. A study of light new physics will be presented elsewhere.
Heavy new physics is parameterized by modifications of the two Wilson coefficients $C_L$ and $C_R$ in the effective Hamiltonian, see equation~\eqref{eq:H_eff}. Heavy new physics can change the values of the branching ratio and the forward-backward asymmetry, as well as their kinematic distributions.

%%%%%%%%%%%%%%%%%%%%%%%%%%%%%%%%%%%%%%
\begin{figure}[tb]
\centering
\includegraphics[width=0.46\textwidth]{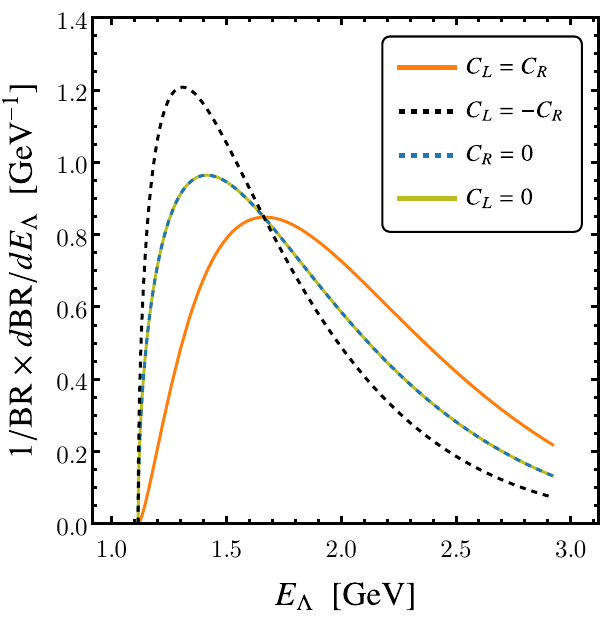} ~~~~ 
\includegraphics[width=0.46\textwidth]{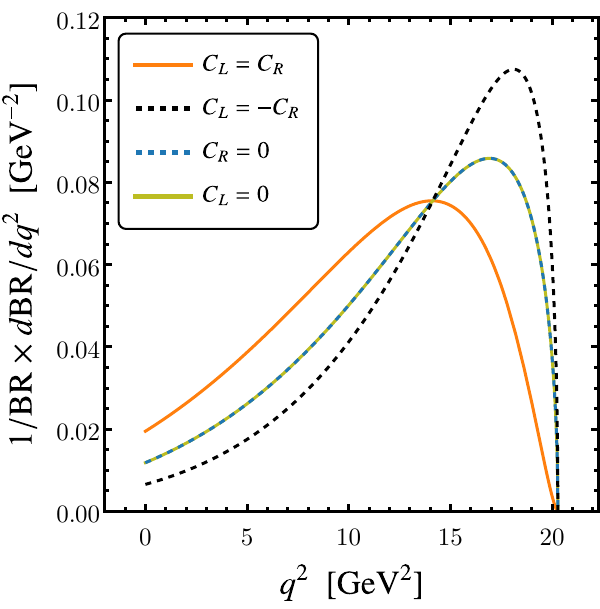}
\caption{The normalized differential branching ratio of $\Lambda_b \to \Lambda \nu \bar \nu$ in various new physics scenarios as a function of the $\Lambda$ energy in the $\Lambda_b$ rest frame $E_\Lambda$ (left) and the di-neutrino invariant mass squared $q^2$ (right). The scenario with $C_R = 0$ coincides with the SM prediction.}
\label{fig:dBR_NP}
\end{figure}
%%%%%%%%%%%%%%%%%%%%%%%%%%%%%%%%%%%%%%
\begin{figure}[tb]
\centering
\includegraphics[width=0.46\textwidth]{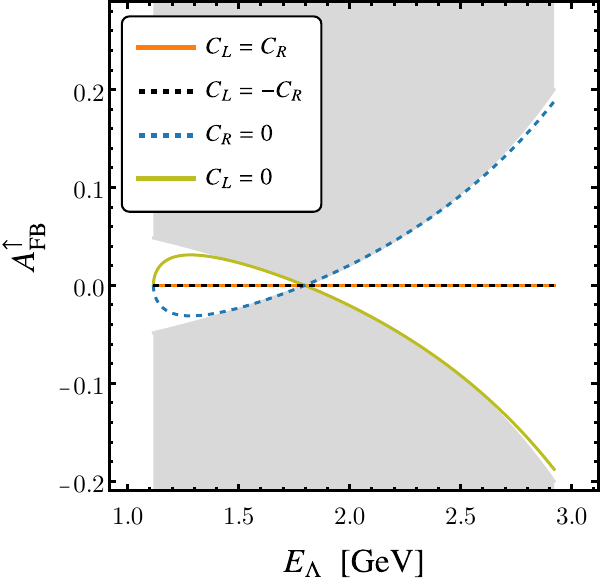} ~~~~
\includegraphics[width=0.46\textwidth]{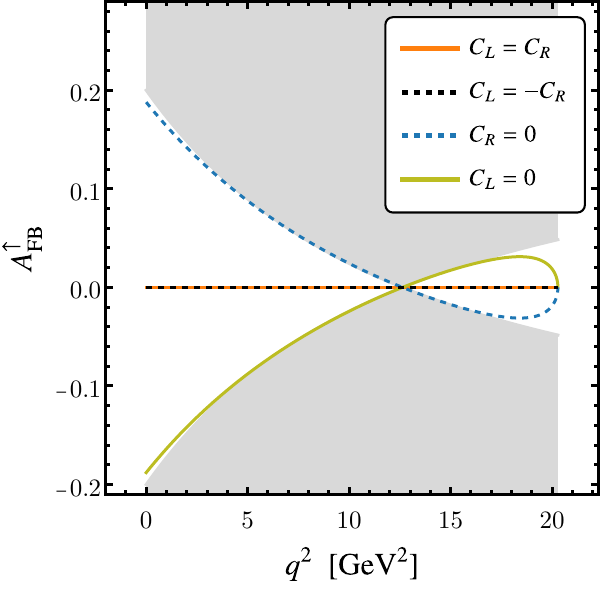}
\caption{The forward-backward asymmetry in $\Lambda_b \to \Lambda \nu \bar \nu$ in various new physics scenarios as a function of the $\Lambda$ energy in the $\Lambda_b$ rest frame $E_\Lambda$ (left) and the di-neutrino invariant mass squared $q^2$ (right). The $\Lambda_b$ polarization is set to its central experimental value $\mathcal P_{\Lambda_b} = -0.40$. The scenario with $C_R = 0$ coincides with the SM prediction.}
\label{fig:AFB_NP}
\end{figure}
%%%%%%%%%%%%%%%%%%%%%%%%%%%%%%%%%%%%%%

In the plots of figure~\ref{fig:dBR_NP}, we show the normalized differential branching ratio as a function of the $\Lambda$ energy (left) and of $q^2$ (right) in four different new physics scenarios. 
Scenarios with only left-handed currents ($C_R = 0$, dashed blue) or only right-handed currents ($C_L = 0$, yellow) result in the same kinematic distributions that coincide the SM prediction. Scenarios with vector currents ($C_L = C_R$, orange) give a slightly harder $E_\Lambda$ spectrum, while the $E_\Lambda$ spectrum for axial-vector currents ($C_L = - C_R$, dashed black) is slightly softer. 

The new physics effects on the forward-backward asymmetry are highly complementary. The plots of figure~\ref{fig:AFB_NP} show the forward-backward asymmetry as a function of the $\Lambda$ energy (left) and of $q^2$ (right) in the same four new physics scenarios. The forward-backward asymmetry maximally distinguishes the two chiral scenarios $C_R = 0$ and $C_L = 0$. The forward-backward asymmetry vanishes for both the vector and axial-vector scenarios.
The gray regions in the plots is theoretically inaccessible. 
In fact, the maximum value of the forward-backward asymmetry that can in principle be reached is given by
\begin{equation}
 \big| A_\text{FB}^\uparrow \big| < \left| \frac{\mathcal P_{\Lambda_b} \mathcal F_{VA}}{2 \sqrt{\mathcal F_V \mathcal F_A}} \right| ~,
\end{equation}
which defines the boundary of the gray region in the plots of figure~\ref{fig:AFB_NP}. In passing, we note that the SM prediction (that coincides with the dashed blue line) is very close to the theoretical maximum of the forward-backward asymmetry. 

Interestingly enough, the zero crossing of the forward-backward asymmetry does not change with new physics, but depends purely on form factor inputs. It is always determined by the condition in equation~\eqref{eq:q2_0}. Our predictions in equation~\eqref{eq:zero_crossing} therefore serve as an interesting experimental cross-check of the form factor calculations.

%%%%%%%%%%%%%%%%%%%%%%%%%
\section{Sensitivity to Heavy New Physics}
\label{sec:NP}
%%%%%%%%%%%%%%%%%%%%%%%%%

After the discussion in the previous section that focused on the qualitative impact that heavy new physics can have on the $\Lambda_b \to \Lambda \nu\bar\nu$ decay, we now determine quantitatively the sensitivity to heavy new physics. 

As discussed above, new physics can modify the value of the branching ratio and the forward-backward asymmetry as well as their kinematic distributions (but not the location of the zero crossing of the forward-backward asymmetry). The new physics sensitivity of the $\Lambda_b \to \Lambda \nu\bar\nu$ observables will depend on the precision of the theory predictions and the uncertainties of the future experimental measurements. The uncertainty of a future $\Lambda_b \to \Lambda \nu\bar\nu$ branching ratio measurement at the FCC-ee was estimated to be around $10\%$ in~\cite{Amhis:2023mpj}. This is below the current theory uncertainty of $\sim 14\%$, see equation~\eqref{eq:SM_prediction}. So far, no estimates exist how well the forward-backward asymmetry could be measured at future $Z$-pole machines.
In our study, we will take into account the current uncertainties on the theory side and will neglect the experimental uncertainties. This illustrates to which extent the current theory precision limits the sensitivity of $\Lambda_b \to \Lambda \nu\bar\nu$. We do expect improved theory predictions by the time the FCC-ee or CEPC would deliver results.

To incorporate new physics to the branching ratio and the forward-backward asymmetry, we find it convenient to normalize the new physics contributions to the Wilson coefficients by the SM Wilson coefficient
\begin{equation}
c_L^\text{NP} = C_L^\text{NP} / C_L^\text{SM} ~,\qquad c_R^\text{NP} = C_R^\text{NP} / C_L^\text{SM} ~.
\end{equation}
The expressions for the branching ratio and the forward-backward asymmetry in the presence of new physics are then 
\begin{eqnarray}
 \text{BR}(\Lambda_b \to \Lambda \nu\bar\nu) &=& \text{BR}(\Lambda_b \to \Lambda \nu\bar\nu)_\text{SM} \Big( r \big| 1 + c_L^\text{NP} + c_R^\text{NP} \big|^2  + (1-r) \big| 1 + c_L^\text{NP} - c_R^\text{NP} \big|^2  \Big) ~, \\
 \langle A_\text{FB}^\uparrow \rangle &=& \frac{  \langle A_\text{FB}^\uparrow \rangle_\text{SM} \Big( \big| 1 + c_L^\text{NP} \big|^2 - \big| c_R^\text{NP} \big|^2 \Big) }{r \big| 1 + c_L^\text{NP} + c_R^\text{NP} \big|^2  + (1-r) \big| 1 + c_L^\text{NP} - c_R^\text{NP} \big|^2} ~,
\end{eqnarray}
where the parameter $r$ is given by
\begin{equation}
r = \frac{\int_{E_\Lambda^\text{min}}^{E_\Lambda^\text{max}} dE_\Lambda E_\Lambda^2 \sqrt{1 - \frac{m_\Lambda^2}{E_\Lambda^2}} ~ \mathcal F_{V}}{\int_{E_\Lambda^\text{min}}^{E_\Lambda^\text{max}} dE_\Lambda E_\Lambda^2 \sqrt{1 - \frac{m_\Lambda^2}{E_\Lambda^2}} ~\big( \mathcal F_V + \mathcal F_A \big) } ~.
\end{equation}
The expressions can be applied for any range of the $\Lambda$ energy or $q^2$, respectively. In particular, we find for the total $q^2$ range and the low and high $q^2$ ranges
\begin{equation}
r_\text{tot} \simeq 0.41 ~,\quad r_\text{low} \simeq 0.55 ~, \quad r_\text{high} \simeq 0.30 ~.
\end{equation}
 
%%%%%%%%%%%%%%%%%%%%%%%%%%%%%%%%%%%%%%
\begin{figure}[tb]
\centering
\includegraphics[width=0.46\textwidth]{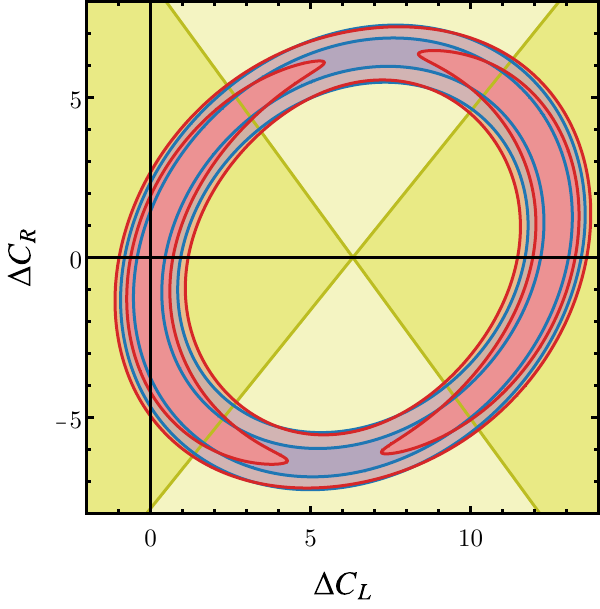} ~~~~~
\includegraphics[width=0.46\textwidth]{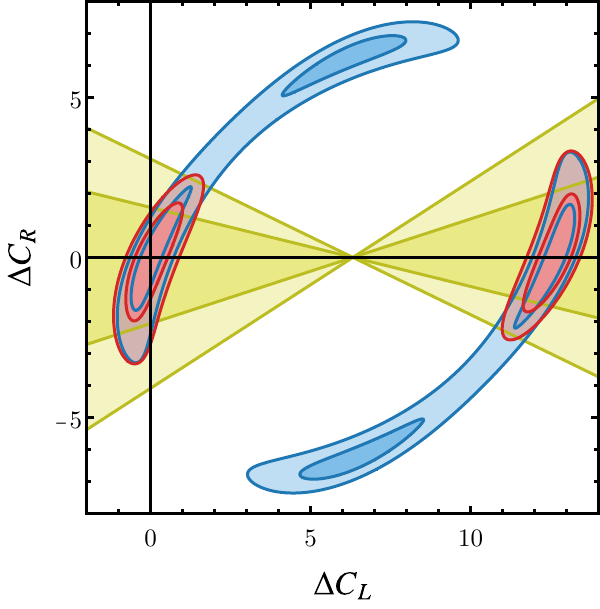}
\caption{The expected sensitivity of $\Lambda_b \to \Lambda \nu\bar\nu$ to new physics parameterized by the Wilson coefficients $C_L^\text{NP}$ and $C_R^\text{NP}$. Left: taking into account a single large $q^2$ bin; Right: splitting the $q^2$ range into two bins, one above and one below the zero crossing of the forward-backward asymmetry. The blue, yellow, and red regions correspond to the constraints from the branching ratio, the forward-backward asymmetry, and their combination, respectively. Dark and light shaded regions indicate $1\sigma$ and $2\sigma$ constraints.}
\label{fig:WCplots}
\end{figure}
%%%%%%%%%%%%%%%%%%%%%%%%%%%%%%%%%%%%%%

In figure~\ref{fig:WCplots}, we show the allowed regions in $C_L^\text{NP}$ vs. $C_R^\text{NP}$ parameter space, assuming that the experimental measurements of the branching ration and the forward-backward asymmetry agree precisely with our SM predictions.
The plot on the left takes into account observables integrated over the full $q^2$ range, while in the plot on the right we include the low and high $q^2$ regions separately, taking into account their correlations. We assume that both $C_L^\text{NP}$ and $C_R^\text{NP}$ are real.

The blue regions correspond to the constraint we anticipate from a future branching ratio measurement at the $1\sigma$ and $2\sigma$ level, while the yellow regions correspond to a measurement of the forward-backward asymmetry. The red regions show the combination of both. More precisely, the dark and light shaded regions correspond to a $\Delta \chi^2$ of 1 and 4, when we consider the branching ratio or the forward backward asymmetry individually (blue and yellow regions), and to a $\Delta \chi^2$ of 2.3 and 6 in the combination (red regions).
As expected, the branching ratio and the forward-backward asymmetry have complementary sensitivity. However, if the forward-backward asymmetry is integrated over the entire $q^2$ range, its constraining power is largely washed out. A much better sensitivity is obtained if the high-$q^2$ and low-$q^2$ regions are taken into account separately, which avoids cancellations in the forward-backward asymmetry. 

In the right plot, we observe two clearly resolved regions of parameter space: one in the vicinity of the SM point $C_L^\text{NP} = C_R^\text{NP} = 0$, and one with $C_L^\text{NP} \simeq - 2 C_L^\text{SM} \simeq 12.6$, $C_R^\text{NP} \simeq 0$. The second region corresponds to a total $b \to s \nu \nu$ amplitude with opposite sign compared to the SM, a scenario that cannot be distinguished from the SM using only $b \to s \nu \nu$ decays. We consider such a scenario contrived and do not discuss it further.

We can translate the allowed region around the point $C_L^\text{NP} = C_R^\text{NP} = 0$ into constraints on the new physics scale, using the parameterization 
\begin{equation}
 \frac{1}{\Lambda_L^2} = \frac{4 G_F}{\sqrt{2}} \frac{\alpha}{4 \pi} |V_{ts}^* V_{tb}| C_L^\text{NP} ~, \quad \frac{1}{\Lambda_R^2} = \frac{4 G_F}{\sqrt{2}} \frac{\alpha}{4 \pi} |V_{ts}^* V_{tb}| C_R^\text{NP} ~.
\end{equation}
We find the following $2\sigma$ limits
\begin{equation}
 \frac{-1}{\big( 45 \,\text{TeV} \big)^2} \lesssim  \frac{1}{\Lambda_L^2} \lesssim \frac{1}{\big( 44 \,\text{TeV} \big)^2} ~, \quad  \frac{-1}{\big( 25 \,\text{TeV} \big)^2} \lesssim  \frac{1}{\Lambda_R^2} \lesssim \frac{1}{\big( 33 \,\text{TeV} \big)^2} ~.
\end{equation}
These new physics scales are in the same ballpark as the scales that are probed by $b \to s \mu\mu$ decays (see e.g.~\cite{Altmannshofer:2017yso, DiLuzio:2017chi}) and by the rare meson decays with neutrinos in the final state $B \to K \nu\bar\nu$ and $B \to K^* \nu \bar\nu$ (see e.g.~\cite{Bause:2023mfe, Allwicher:2023xba}).

%%%%%%%%%%%%%%%%%%%%%%%%%
\section{Differential Branching Ratio in the Lab Frame}
\label{sec:lab_frame}
%%%%%%%%%%%%%%%%%%%%%%%%%

As we have seen in the previous section, a measurement of the forward-backward asymmetry provides interesting sensitivity to new physics that is complementary to measurements of the branching ratio. 
We note that a measurement of $A_\text{FB}^\uparrow$ requires access to the angle between the flight direction of the $\Lambda$ in the $\Lambda_b$ rest frame and the flight direction of the $\Lambda_b$ in the lab frame. Reconstructing the $\Lambda_b$ rest frame might be challenging due to the two neutrinos in the decay\footnote{One could for example estimate the di-neutrino energy and momentum using the missing energy and momentum in the hemisphere containing the $\Lambda_b$.}, and we therefore explore to which extent information about the forward-backward asymmetry (or, equivalently, information about the chirality structure of the $b \to s \nu \nu$ transition) can be extracted from lab frame observables alone. The approach is inspired by methods to determine the $\Lambda_b$ polarization at LEP, using semi-leptonic decays~\cite{Mannel:1991bs, Mele:1992kh, Bonvicini:1994mr}. 

For the following, it will be useful to introduce the shorthand notation
\begin{equation}
\hat \beta_{\Lambda_b} =  \sqrt{1 - \frac{m_{\Lambda_b}^2}{\hat E_{\Lambda_b}^2}} ~, \quad \hat \beta_\Lambda = \sqrt{ 1 - \frac{m_\Lambda^2}{\hat E_\Lambda^2} } ~, \quad \beta_\Lambda = \sqrt{ 1 - \frac{m_\Lambda^2}{E_\Lambda^2} } ~,
\end{equation}
where $\hat \beta_{\Lambda_b}$ is the velocity of the $\Lambda_b$ in the lab frame, $\hat \beta_\Lambda$ the velocity of the $\Lambda$ in the lab frame, and $\beta_\Lambda$ the velocity of the $\Lambda$ in the $\Lambda_b$ restframe, respectively.

The relevant lab frame observable is the differential decay rate as a function of the $\Lambda$ energy in the lab frame that we denote with $\hat E_\Lambda$. We find
\begin{equation} \label{eq:dGamma_dhatE}
    \frac{d\text{BR}(\Lambda_b \to \Lambda \nu\bar\nu)}{d \hat E_\Lambda} = \int_{E_\Lambda^\text{min}}^{E_\Lambda^\text{max}} \frac{dE_\Lambda}{E_\Lambda} \frac{m_{\Lambda_b}}{\hat E_{\Lambda_b}} \frac{1}{\hat \beta_{\Lambda_b} \beta_\Lambda} \frac{d \text{BR}(\Lambda_b \to \Lambda \nu\bar\nu)}{dE_\Lambda d\cos\theta_\Lambda} ~,
\end{equation}
where $\hat E_{\Lambda_b}$ is the energy of the $\Lambda_b$ in the lab frame, and in the expression for the differential branching ratio in the $\Lambda_b$ restframe~\eqref{eq:dBR_dE_dcos}, one needs to replace
\begin{equation}
    \cos\theta_\Lambda = \frac{1}{\hat \beta_{\Lambda_b} \beta_\Lambda} \left( \frac{\hat E_\Lambda}{E_\Lambda} \frac{m_{\Lambda_b}}{\hat E_{\Lambda_b}} - 1 \right) ~.
\end{equation}
The lower and upper integration boundaries in~\eqref{eq:dGamma_dhatE} are given by
\begin{eqnarray}
    E_\Lambda^\text{min} &=& \frac{\hat E_\Lambda  \hat E_{\Lambda_b}}{m_{\Lambda_b}} \left(1 - \hat \beta_{\Lambda_b} \hat \beta_\Lambda \right) ~, \\
    E_\Lambda^\text{max} &=& \text{Min}\left\{ \frac{m_{\Lambda_b}}{2} \left(1 + x_\Lambda^2 \right) ,~ \frac{\hat E_\Lambda  \hat E_{\Lambda_b}}{m_{\Lambda_b}} \left(1 + \hat \beta_{\Lambda_b} \hat \beta_\Lambda \right) \right\} ~.
\end{eqnarray}
The physical range of the $\Lambda$ energy in the lab frame is 

{\everymath{\displaystyle} 
\begin{eqnarray}
    \hat E_\Lambda^\text{min} &=& \begin{cases} m_\Lambda \quad\qquad\qquad\qquad\qquad\qquad\qquad\qquad\quad ~ \text{if} ~~  \hat \beta_{\Lambda_b} < \frac{1-x_\Lambda^2}{1+x_\Lambda^2} ~, \\ \frac{\hat{E}_{\Lambda_b}}{2} \left( 1 + x_\Lambda^2 - \left( 1 - x_\Lambda^2 \right)  \hat \beta_{\Lambda_b} \right)  \quad\qquad\qquad \text{if} ~~ \hat \beta_{\Lambda_b} > \frac{1-x_\Lambda^2}{1+x_\Lambda^2} ~, \end{cases} \\[12pt]
    \hat E_\Lambda^\text{max} &=& \frac{\hat{E}_{\Lambda_b}}{2} \left( 1 + x_\Lambda^2 + \left( 1 - x_\Lambda^2 \right)  \hat \beta_{\Lambda_b} \right) ~.
\end{eqnarray}}

The $\hat E_\Lambda$ spectrum depends on $\hat E_{\Lambda_b}$, the energy of the decaying $\Lambda_b$. The energy spectrum of $B$ mesons produced on the $Z$-pole has been measured at LEP~\cite{SLD:1997ulk}. The average energy is around $70\%$ of half the center of mass energy. As argued in~\cite{Bonvicini:1994mr} the average energies of mesons and baryons should be very similar and we therefore expect an average energy of $\Lambda_b$ baryons of around (30-35)\,GeV, however, with a distribution that covers also much smaller and much larger values. 

%%%%%%%%%%%%%%%%%%%%%%%%%%%%%%%%%%%%%%
\begin{figure}[tb]
\centering
\includegraphics[width=0.46\textwidth]{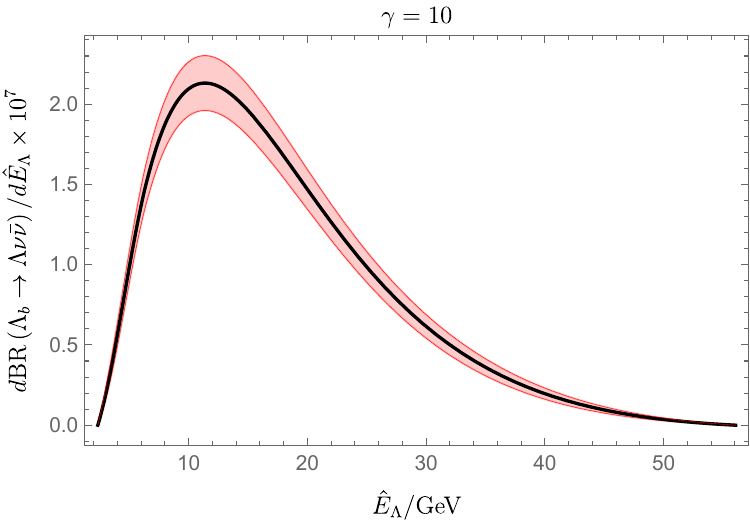}
\includegraphics[width=0.46\textwidth]{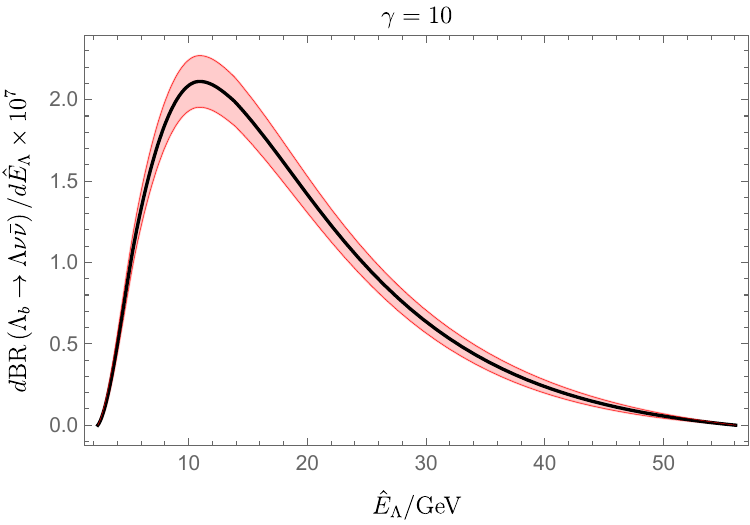}
\includegraphics[width=0.46\textwidth]{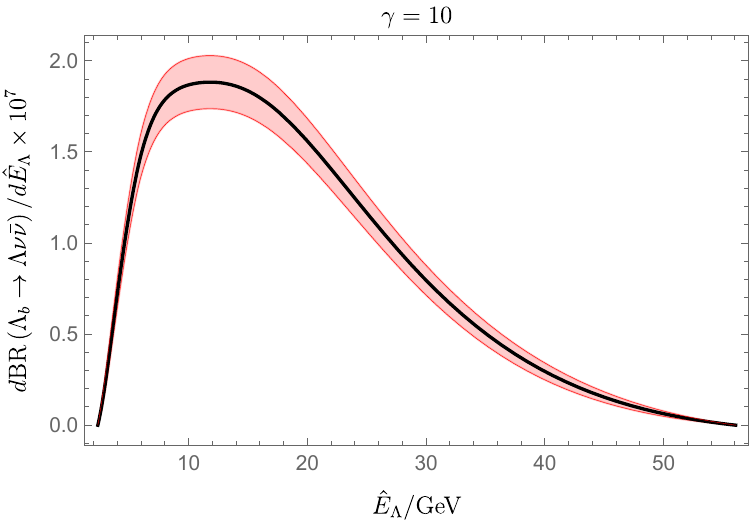}
\includegraphics[width=0.46\textwidth]{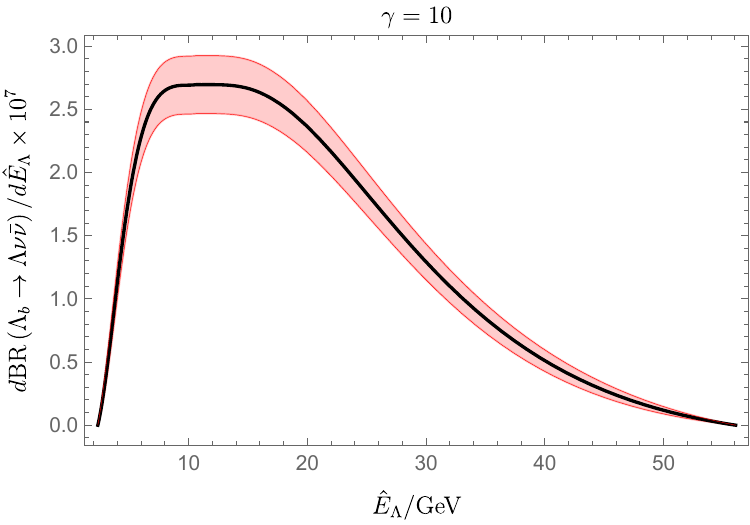}
\caption{The differential branching ratio $\Lambda_b \to \Lambda \nu\bar\nu$ as a function of the $\Lambda$ energy in the lab frame. For illustration, we set the $\Lambda_b$ energy to $\hat E_{\Lambda_b} = 10 ~m_{\Lambda_b} \simeq 56$\,GeV. We show a few benchmark cases: $C_L = C_L^\text{SM}$, $C_R = 0$ (top left), $C_L = 0$, $C_R = C_L^\text{SM}$ (top right), $C_L=2C_R=C_L^\text{SM}$ (bottom left), and $C_L=C_R=C_L^\text{SM}$ (bottom right). The shaded band corresponds to the $1\sigma$ theory uncertainty on the branching ratio.}
\label{fig:E_Lambda_lab}
\end{figure}
%%%%%%%%%%%%%%%%%%%%%%%%%%%%%%%%%%%%%%

We find that the differential branching ratio \eqref{eq:dGamma_dhatE} scales to a very good approximation with the $\Lambda_b$ energy, i.e. its shape is approximately invariant if plotted as a function of $\hat E_\Lambda / \hat E_{\Lambda_b}$. For illustration, we show in figure~\ref{fig:E_Lambda_lab} the differential branching ratio as a function of the $\Lambda$ energy in the lab frame for a fixed $\hat E_{\Lambda_b} = 10 ~m_{\Lambda_b} \simeq 56$\,GeV. The four panels correspond to four benchmark cases for the Wilson coefficients: $C_L = C_L^\text{SM}$, $C_R = 0$ (top left), $C_L = 0$, $C_R = C_L^\text{SM}$ (top right), $C_L=2C_R=C_L^\text{SM}$ (bottom left), and $C_L=C_R=C_L^\text{SM}$ (bottom right). The peak of the distributions is at a $\Lambda$ energy of around $\hat E_\Lambda \simeq 0.2 \times \hat E_{\Lambda_b}$ in all cases. The distribution is much broader if both $C_L$ and $C_R$ are present.

The distributions for purely left-chiral interactions (top left) and purely right-chiral interactions (top right) are very similar. This result indicates that the effect of the forward backward asymmetry (that maximally distinguishes between the left-chiral and right-chiral scenarios, see figure~\ref{fig:AFB_NP}) is largely washed out in the lab frame. Dedicated sensitivity studies are required to determine to which extent a lab frame analysis could recover information about the chirality structure, or if a $\Lambda_b$ rest frame analysis can provide much better sensitivity.

%%%%%%%%%%%%%%%%%%%%%%%%%
\section{Conclusions}
\label{sec:conclusions}
%%%%%%%%%%%%%%%%%%%%%%%%%

In this paper, we explored the new physics sensitivity of the rare decay $\Lambda_b \to \Lambda \nu \bar\nu$. The decay has never been observed but can be accessed at future $Z$-pole machines like FCC-ee and CEPC. In contrast to the mesonic counterparts ($B \to K \nu \bar\nu$, $B \to K^* \nu\bar\nu$, and $B_s \to \phi \nu\bar\nu$), $\Lambda_b$ baryons produced in $Z$ decays are longitudinally polarized, which offers complementary novel observables to test the Standard Model. 

The kinematics of the $\Lambda$ baryon in the final state depends on two independent variables, its energy and the angle between its momentum and the spin of the $\Lambda_b$. We find that the corresponding angular decay distribution is characterized by a forward-backward asymmetry, $A_\text{FB}^\uparrow$, that is proportional to the longitudinal polarization of the $\Lambda_b$. We give state-of-the-art predictions for the integrated branching ratio, $\text{BR}(\Lambda_b \to \Lambda \nu\bar\nu)$, and the forward-backward asymmetry, $A_\text{FB}^\uparrow$, in the Standard Model. The uncertainties are dominated by the current knowledge of $\Lambda_b \to \Lambda$ form factors, providing continued motivation to improve baryon form factor calculations on the lattice. We find that the forward-backward asymmetry has a zero crossing at a specific value of the $\Lambda$ energy, $E_\Lambda$ (or equivalently of the di-neutrino invariant mass, $q^2$). Interestingly, the location of the zero-crossing is independent of new physics and therefore can be used as an experimental test of form factor calculations. On the other hand, the $q^2$ shapes of the differential branching ratio and the forward-backward asymmetry do depend on new physics.

Parameterizing heavy new physics in $\Lambda_b \to \Lambda \nu\bar\nu$ by an effective Hamiltonian, we determine the expected new physics sensitivity of future precision measurements of $\Lambda_b \to \Lambda \nu\bar\nu$ on the $Z$-pole. We find that splitting the $q^2$ region into two bins, one above the $A_\text{FB}^\uparrow$ zero crossing and one below, greatly enhances the new physics sensitivity, and allows one to break degeneracies in the new physics parameter space. We find that new physics scales of $\Lambda_\text{NP} \sim (25 - 45)$\,TeV can be probed. This is comparable to the scales that can be probed with other $ b\to s \nu\bar\nu$ decays and with $b \to s \ell^+ \ell^-$ decays. It would be interesting to extend our study to the $\Lambda_b \to \Lambda \ell^+ \ell^-$ decays on the $Z$-pole. Also in that case we expect that the longitudinal polarization of the $\Lambda_b$ gives novel probes of new physics. 

An experimental measurement of the forward-backward asymmetry requires reconstruction of the $\Lambda_b$ rest frame. Because of the presence of the two neutrinos in the final state, this may be challenging. We therefore also explored to which extent information about the forward-backward asymmetry can be accessed from lab frame observables.
We find that different values of the forward-backward asymmetry leave only a very mild imprint on the energy distribution of the $\Lambda$ in the lab frame. More detailed studies are required to determine if a lab frame analysis or a $\Lambda_b$ rest frame analysis would have better sensitivities.

%%%%%%%%%%%%%%%%%%%%%%%%%%%%%%%%
\section*{Acknowledgements}
%%%%%%%%%%%%%%%%%%%%%%%%%%%%%%%%

We thank Howard Haber and Jason Nielsen for useful discussions. The research of WA, SAG, and KT is supported by the U.S. Department of Energy grant number DE-SC0010107.

\begin{appendix}
%%%%%%%%%%%%%%%%%%%%%%%%%
\section{Form Factors}
\label{app:ff}
%%%%%%%%%%%%%%%%%%%%%%%%%

In this appendix we give details about our implementation of the $\Lambda_b \to \Lambda$ form factors. We follow~\cite{Detmold:2016pkz} and parameterize the form factors using a $z$-expansion
\begin{equation}
f(q^2) = \frac{1}{1- q^2 / (m^f_\text{pole})^2} \sum_n  a_n^f z^n ~, \qquad z = \frac{\sqrt{t_+ - q^2} - \sqrt{t_+ - t_0}}{\sqrt{t_+ - q^2} + \sqrt{t_+ - t_0}} ~.
\end{equation}
The parameters $t_0$ and $t_+$ are chosen such that the kinematic endpoint $q^2_\text{max}$ is mapped to $z = 0$, and $q^2$ values above the $BK$ threshold are mapped onto the unit circle in the complex $z$ plane
\begin{equation}
t_0 = ( m_{\Lambda_b} - m_\Lambda)^2 ~, \qquad t_+ = (m_B + m_K)^2~.
\end{equation}
The poles that correspond to the lowest relevant $B_s$ mesons are factored out explicitly. The corresponding masses $m_\text{pole}^f$ and the expansion coefficients $a_n^f$ are taken from~\cite{Detmold:2016pkz}.

%%%%%%%%%%%%%%%%%%%%%%%%%%%%%%%%%%%%%%
\begin{figure}[tb]
\centering
\includegraphics[width=0.46\textwidth]{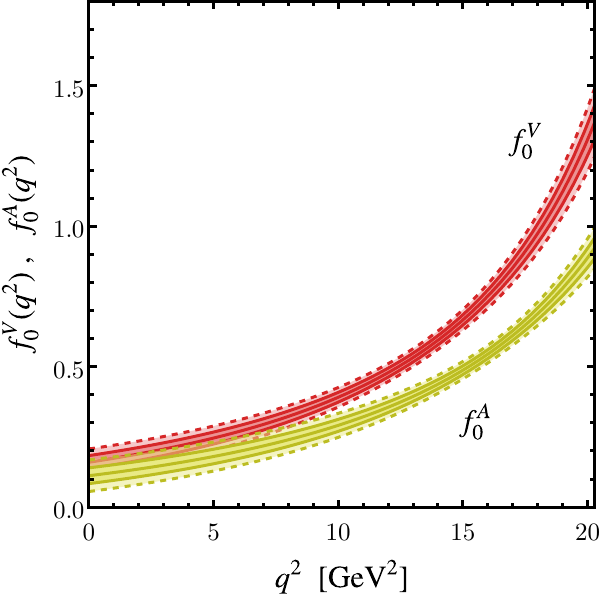} ~~~~ \includegraphics[width=0.46\textwidth]{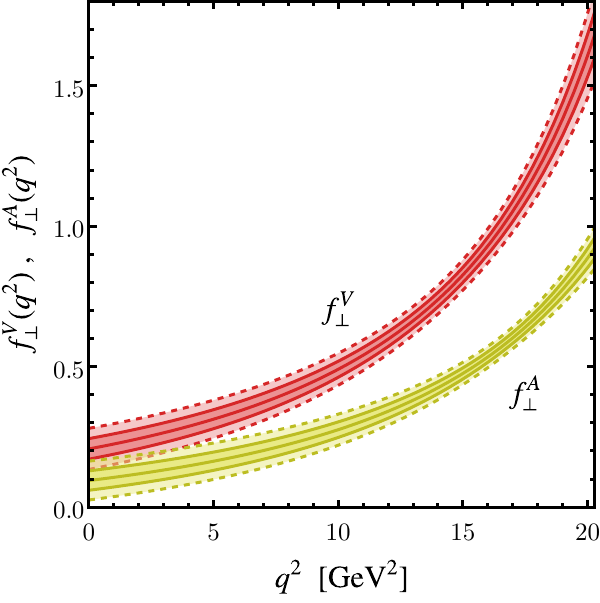} 
\caption{The relevant $\Lambda_b \to \Lambda$ form factors as a function of $q^2$. The colored bands correspond to the $1\sigma$ and $2\sigma$ uncertainties.}
\label{fig:ff_LbtoL}
\end{figure}
%%%%%%%%%%%%%%%%%%%%%%%%%%%%%%%%%%%%%%

To obtain the form factor uncertainties, we follow the procedure recommended in~\cite{Detmold:2016pkz} that takes into account the difference in the results based on $z$-expansions up to order $n=1$ and $n=2$. Figure~\ref{fig:ff_LbtoL} shows all the form factors relevant for our analysis with their $1\sigma$ and $2\sigma$ uncertainties.

\end{appendix}

%%%%%%%%%%%%%%%%%%%%%%%%%%%
\bibliographystyle{jhep}
\bibliography{refs} 
%%%%%%%%%%%%%%%%%%%%%%%%%%%

\end{document}